\documentclass[epj,referee]{svjour}
\input{epsf.sty}
\begin{document}

\title{Long-range spin-pairing order and spin defects in
quantum spin-$\frac{1}{2}$ ladders}

\author{M.A. Garcia-Bach}

\institute{Departament de F\'{\i}sica Fonamental, Facultat de
F\'{\i}sica, Universitat de Barcelona, Diagonal 647,
E-08028 Barcelona, Catalunya, Spain. \email{angels@hermes.ffn.ub.es}}

\date{Received:  / Revised version: \today }

\abstract{For $w$-legged antiferromagnetic spin-1/2 Heisenberg
ladders, a long-range spin pairing order can be identified which
enables the separation of the space spanned by finite-range
(covalent) valence-bond configurations into $w+1$ subspaces.
Since every subspace has an equivalent counter subspace connected by
translational symmetry, twofold degeneracy, breaking translational
symmetry is found except for the subspace where the ground state of
$w$=even belongs to.
In terms of energy ordering, (non)degeneracy and the discontinuities
introduced in the long-range spin pairing order by topological spin
defects, the differences between even and odd ladders are explained
in a general and systematic way. 
 }

\maketitle

\newpage
\section{Introduction}
\label{intro}

The discovery, about a decade ago, of high-Tc superconductivity
\cite{Bednorz} in lightly doped ``two dimensional'' antiferromagnets
and materials (initially) supposed to contain coupled spin chains
\cite{Johnston,Hiroi,Takano}, have generated a renewed interest on
low dimensional quantum spin-$\frac{1}{2}$ systems.  One of the
concerns is the non-smooth crossover from one-dimensional to 
two-dimensional systems (see, for instance, Ref.\ \cite{Dagotto}
and references therein).  This fact has also been pointed out earlier
in Refs.\ \cite{Seitz,strips1,strips2,Tanaka-87} for different sets
of long polymeric strips with graphite as the final member of these
series, paralleling that of the square-lattice family.  Both
theoretical and experimental studies
\cite{Dagotto,Rojo,Frischmuth,Greven} suggest that the nature of
antiferromagnetic spin-$\frac{1}{2}$ ladders with $w$=even legs
differs from that of $w$=odd ladders.  For instance, $w$=even ladders
are gapped systems, the gap vanishing exponentially with $w$, while
$w$=odd ladders display characteristics similar to one-dimensional
spin-$\frac{1}{2}$ systems, namely they are gapless, with a doubly
degenerate ground state, breaking translational symmetry
\cite{Rojo,LSM,Yamanaka}.  Furthermore, spin defects are confined in
ladders with $w$=even but they are not if $w$=odd.  Numerical results
\cite{Manousakis} indicate that, in the infinite limit, the ground
state of the two dimensional system, towards $w$=even and $w$=odd
series must converge to, has long-range antiferromagnetic order and
gapless excitations.

In this paper we will consider antiferromagnetic quantum
spin-$\frac{1}{2}$ ladders with $w$ legs, (even) $L\rightarrow\infty$
rungs, free boundary conditions in the inter-chain direction, and
translational symmetry in the chain direction.  It is assumed that
the Hamiltonian appropriate to describe these systems contains only
short-range interactions preserving the total spin of the
system.  At half filling, we assume that the Hamiltonian which
governs the lowest-lying region of the spectrum is the
spin-$\frac{1}{2}$ antiferromagnetic Heisenberg Hamiltonian,
\begin{equation}
H = \sum_{ni,mj}J_{ni,mj}{\bf S}_{ni} \cdot {\bf S}_{mj}.
\end{equation}
where ${\bf S}_{ni}$ is the spin operator for spin on site $ni$, $n$
indicating the rung and $i$ the leg, and the $J_{ni,mj}$ are the
exchange-coupling parameters.  The $J_{ni,mj}$ are assumed to
decrease very rapidly with distance, the nearest-neighbour Heisenberg
Hamiltonian with isotropic $J$ being the dominant part of $H$.

Since the ground state of such a Hamiltonian for a bipartite system
with equal number of sites in the two parts is known to be a singlet
\cite{Lieb-Mattis}, resonating-valence-bond-type wave functions are
defined in the space spanned by $M$-range (covalent) valence-bond
(VB) configurations, with arbitraryly large but finite $M$.  We
refer to this space as {\em model space\/}, ${\cal H}^w$. The
reasonableness of ${\cal H}^w$ is based on the fact that the
dimer-covering configurations (or Kekul\'e structures
\cite{Pauling}, as have always been termed in Resonance Theory) are
the lowest-lying monoconfigurational singlets.  Thus, they provide a
good zero-order picture.  Then, on applying the Hamiltonian $H$, it
can be noticed that the nearness of spin pairing tends to be
preserved.  When $H$ is restricted to the isotropic nearest-neighbour
spin-$\frac{1}{2}$ Heisenberg Hamiltonian the shorter-range RVB
picture should apply best for small even $w$, while $w$=odd or wide
even $w$ ladders are expected to require long-range RVB pictures
\cite{PRB-91,White}.  For instance, $M$-range RVB pictures neglect
corrections lying higher than the $M$ order in Perturbation Theory
and have to be considered with caution.  Nevertheless, additional
terms in the Hamiltonian, as frustration, are expected to stabilise
the finite-range RVB wave functions with respect to other
N\'eel-based {\em ans\"atze\/} (see \cite{PhysLet82} and references
therein).  In addition, there exist finite-range Heisenberg models
for which short-range Kekul\'{e} structures are exact ground states
and also short-range RVB {\em ans\"atze\/} certainly apply for
so-called ``bond-dimer'' models (see, for instance,
\cite{Majumdar,van-den-Broek,SShastry-S,Klein-82,Miyahara}).

In order to separate the model space ${\cal H}^w$ into non-mixing
different subspaces, several attempts have been made to find
associated topological quantum numbers.  For instance, the
occurrence of a topological long-range order (LRO) was first
discussed \cite{PRB-79} to rationalise the ground-state instability
to bond alternation in spin-1/2 linear Heisenberg chains.
Simultaneously, this LRO has also been discussed in the context of
applications to conjugated hydrocarbons
\cite{Seitz,strips1,strips2,Klein-79}.  Latter, Klein {\it et al.\/}
\cite{Klein-87} and independently Thouless \cite{Thouless}
introduced the {\it gap\/} or {\it resonance parity\/}, and Kivelson
{\it et al.\/} \cite{Kivelson} and Sutherland \cite{Sutherland},
defined the {\it winding number}.  These numbers allow the separation
of the short-range VB states for odd-width strips in two subsets
leading to degeneracy \cite{Bonesteel,Klein-91}.  The relation
between topological LRO and winding numbers is given in
Ref.\ \cite{Klein-91}.  Also, in Ref.\ \cite{Klein-91}, a {\em
resonance quantum number\/}, $D_n$, which specifies the local (at
boundary n) array of singlets, has been defined for VB systems with
bipartitioning conditions.  Still, arguments based on a topological
LRO have been applied to the qualitative analysis of distortions,
excitations and their coupling for square-lattice strips
\cite{PRB-91} and, more quantitatively, to different polymers
\cite{pace,polimers}.  More recently, simple topological effects
in short-range RVB were also predicated in
Refs.\ \cite{White,White-97} for coupled Heisenberg Chains, based on
numerical results from density matrix renormalization group (DMRG)
techniques on clusters.

Our purpose in this paper is to show that for antiferromagnetic
quantum spin-$\frac{1}{2}$ ladders a long-range spin-pairing order
(LR-SPO) associated with the resonance quantum number $D_n$ can be
defined.  This LR-SPO allows to separate the model space
${\cal H}^w$ into $w+1$ subspaces.  Configurations belonging to
mutually different subspaces should differ repeatedly on each of the
L rungs of the ladder.  Then, they are asymptotically orthogonal, and
never mix by applying a few-particle operator. 

The energy ordering among the lowest-lying state in every subspace
is estimate by the dimer-covering-{\em counting\/} approximation
\cite{Seitz,Klein-86,Zivkovic}.  Counting the dimer-covering
configurations has been achieved by a transfer-matrix technique (see,
i.e., Refs.\ \cite{pace,polimers} and references therein).  Also,
since every subspace has an equivalent counter subspace connected by
translational symmetry, twofold degeneracy is naturally obtained
except for the subspace including the ground state of $w$=even
ladders, irrespective of the details of any Hamiltonian preserving
translational symmetry.  Furthermore, in the present paper it is
shown that a topological spin defect introduces a discontinuity in
the LR-SPO, except for the ground state of $w$=even ladders.  Then,
understanding energy ordering, degeneracy, and the discontinuities
introduced in the LR-SPO by topological spin defects, allow a
general and systematic explanation of the differences between even
and odd ladders.

This paper is organised as follows:  In Sec.\ \ref{sec:SPO} we show
that the (covalent) VB configurations have a LR-SPO, which allows the
separation of the model space in different (asymptotically
orthogonal and non-interacting) subspaces. In
Sec.\ \ref{sec:counting} the energy of the lowest lying state
in every subspace is estimated within the
dimer-covering-{\em counting\/} approximation.  In
Sec.\ \ref{sec:defects} we obtain the discontinuity in the LR-SPO
associated to the presence of a topological spin defect.  In
Sec.\ \ref{sec:results} the results are presented and discussed.
Finally, the conclusions are collected in
Sec.\ \ref{sec:conclusions}.

\section{Singlets and long-range spin-pairing order}
\label{sec:SPO}

Quantum spin-$\frac{1}{2}$ ladders with $N=w \times L$ sites, with
(even) $L\rightarrow\infty$ and free boundary conditions along the
interchain direction, are bipartite system with a singlet ground
state.  Therefore, the ground state can be written as a weighted
superposition of a non-orthogonal complete basis set of singlets,
$\mid s_i\rangle$, $i=1$ to $d_N$,
\begin{equation}
d_N=\frac{N!}{(N/2 +1)!(N/2)!}.
\end{equation}
In a bipartite system, sublattices $A$ and $B$ (starred) can be
identified and a set of $d_N$ independent singlets can be
constructed by pairing to a singlet each of the $N/2$ spins in $A$
to a spin in $B$\@.  We represent one of these spin-pairings (SP) by
an arrow from the site in the sublattice $A$ to its partner in
$B$\@ (see, for instance, Fig.\ \ref{fig:sing}, where a complete set
of linearly-independent singlets for six-site systems are
represented).  Overlap, $\langle s_i \mid s_j \rangle$, and matrix
elements, $\langle s_i\mid H \mid s_j \rangle$ can be evaluated using
the Pauling's \cite{Pauling,Pauling-2} superposition rules.

For the sake of simplicity, we first introduce the LR-SPO of VB
configurations defining the (local, at boundary $n$) resonance
quantum numbers, $D_n$, when boundaries are chosen to run parallel
to rungs, and the model space is restricted to the dimer-covering
approximation.  Later we show that the inclusion of longer pairings
and/or using more general boundaries does not spoil this
LR-SPO\@.  The only effect of selecting boundaries of a different
shape is changing the origin of the LR-SPO parameter.  Finally, in
this section, we discuss the consequences of the LR-SPO on the
eigenstates of $H$ and their degeneracy.

\subsection{Dimer-covering model-space approximation}
\label{ssec:Kekule}

For any Kekul\'e structure, let us define $P^+_n$ ($P^-_n$) as the
number of arrows pointing to the right (left) across a boundary,
$f_n$, lying midway between rungs $n$ and $n+1$ (see
Fig.\ \ref{fig:LRO-K}), and $I_n$ as the number of SP with both
sites in the rung $n$,
\begin{equation}
I_n = 0, 1, \dots \frac{w-b}{2},
\end{equation}
where
\begin{equation}
b \equiv \left\{\begin{array}{lrl}
0, & w = & {\rm  even}, \\
1, & w = & {\rm  odd}.
\end{array} \right.
\end{equation}
If $w^A_n$ ($w^B_n$) is the number of sites belonging to the
intersection of rung $n$ and sublattice $A$ ($B$), it can be easily
seen that
\begin{eqnarray}
w^A_n = P^-_{n-1} + P^+_n +  I_n,  \nonumber  \\
w^B_n = P^+_{n-1} + P^-_n +  I_n.  \label{eqn:valans}
\end{eqnarray}
Choosing $A$ and $B$ sublattices according to
\begin{equation}
w^A_0 - w^B_0 = b,
\end{equation}
it can be written
\begin{eqnarray}
w^A_n = \frac{1}{2}\left[ w + (-1)^n b\right] , \nonumber  \\
w^B_n = \frac{1}{2}\left[ w - (-1)^n b\right] . \label{eqn:w_ns}
\end{eqnarray}
Substracting Eqs.\ (\ref{eqn:valans}) and using
Eqs.\ (\ref{eqn:w_ns})
\begin{equation}
P^-_{n-1} - P^+_{n-1} + P^+_n - P^-_n = w^A_n - w^B_n = (-1)^n b.
\end{equation}
Defining the resonance quantum number $D_n$ at boundary $f_n$ as
\begin{equation}
D_n \equiv P^+_n - P^-_n,
\end{equation}
we obtain
\begin{equation}
D_n = D_{n-1} + (-1)^n b.  \label{eqn:rec-Ds}
\end{equation}
Then a SPO parameter $D \equiv D_0$ can be associated to any
dimer-covering configuration, so that
\begin{equation}
D_n = D - \frac{1}{2}\left[ 1 - (-1)^n \right] b.  \label{eqn:DnD0}
\end{equation}
Since
\begin{eqnarray}
P^+_n = 0, 1, \dots w^A_n, \nonumber \\
P^-_n = 0, 1, \dots w^B_n,
\end{eqnarray}
$D$ can take $w+1$ different values,
\begin{equation}
\label{eqn:Des}
D = \frac{w+b}{2}, \frac{w+b-2}{2}, \dots , \frac{b-w}{2},
\end{equation}
and the (dimer-covering) model space can be partitioned in $w+1$
subspaces, ${\cal H}^w_D$, according to the value of $D$\@.

\subsection{General formulation of the LR-SPO}
\label{ssec:general}

The LR-SPO and $D$ introduced above have been related, for the sake
of simplicity, exclusively to dimer-covering configurations
and to boundaries running parallel to rungs.  Now we remove these
restrictions.  We will see that the shape of the boundaries
limiting fragments of the ladder or the inclusion of long-range
spin-pairings is irrelevant and that the LR-SPO can still be defined.

The dimer-covering model space is not invariant under the Hamiltonian
operator. For instance, the XY terms, $S_{ni}^{\pm} S_{mj}^{\mp}$, of
the nearest-neighbour Heisenberg Hamiltonian acting on a Kekul\'e
structure yield singlets with SP between sites up to 3 bonds apart
(see Fig.\ \ref{fig:recoup}).  Then, as a first step,
linearly independent singlets with 3-bond-range (3BR) SP should be
incorporated into ${\cal H}^w$ to go beyond the dimer-covering
approximation.  These 3BR-SP states allow sites in $A$-sublattice
to be SP to sites in $B$-sublattice no more than 3 bonds apart.
These states can be directly generated by the ``re-coupling''
\cite{pace,polimers} of two simply neighbouring dimers,
i.e.\ unlinked bond-pairs with one and only one site in a pair being
a nearest neighbour to a site in the other pair.  It is worth noting
that these re-couplings satisfy the non-crossing rules.
Then, the 3BR-SP model subspace incorporates any singlet obtained
from a dimer-covering singlet allowing an arbitrary number of
unlinked re-couplings of two simply neighbouring bond-pairs.  Still
longer-range model spaces can be obtained allowing 5BR-SP, 7BR-SP,
$\dots$, to be included in ${\cal H}^w$.  Nevertheless, singlets
with very long bond-range SP should contribute less, so a reasonable
model space will be that including singlets with SP up to $M$ bonds
apart, $M$ not necessarily small.

Let us now also allow the boundary $f^g_n$ to be a line running from
one side of the ladder to the other side, with $n1$ being the first
site to the left of $f^g_n$ in leg 1.  We assume that $f^g_n$ can go
up and down, but it is self-avoiding and is not hitting any site.
Thus, $f^g_n$ must unambiguously break up the ladder in two regions:
left region, $L_n$, and right region, $R_n$ (see
Fig.\ \ref{fig:fgn}).  Therefore, two non-intersecting boundaries,
$f^g_n$ and $f^g_{m}$, $n<m$, define a fragment, $F^g_{n,m}$, as the intersection of $R_n$ and $L_m$. 

We define $P^{g+}_n$ ($P^{g-}_n$) as the number of arrows penetrating
the boundary $f^g_n$ with the arrowhead in the $R_n$ ($L_n$) region.
$I^g_{n,m}$ is the number of arrows with both ends in $F^g_{n,m}$.
$l^g_{n,m}$ ($r^g_{n,m}$) is the number of arrows starting in $R_m$
($L_n$) and with the arrowhead in $L_n$ ($R_m$), i.e.\ with no
partner belonging to $F^g_{n,m}$.  Finally, $F^{gA}_{n,m}$
($F^{gB}_{n,m}$) is the number of sites belonging to the
intersection of $F^g_{n,m}$ and sublattice $A$ ($B$).  Then
\begin{eqnarray}
P^{g-}_n - l^g_{n,n+p} + P^{g+}_{n+p} - r^g_{n,n+p} + I^g_{n,n+p}
& = & F^{gA}_{n,n+p},  \nonumber  \\
P^{g+}_n - r^g_{n,n+p} + P^{g-}_{n+p} - l^g_{n,n+p} + I^g_{n,n+p}
& = & F^{gB}_{n,n+p}.  \label{eqn:valans2}
\end{eqnarray}
Subtracting these two equations we obtain
\begin{equation}
D^g_{n+p} - D^g_n = F^{gA}_{n,n+p} - F^{gB}_{n,n+p},
\label{eqn:DDgeneral}
\end{equation}
with
\begin{equation}
D^g_n \equiv P^{g+}_n - P^{g-}_n,
\end{equation}

Let us analyze $F^g_{n,n+p}$ leg by leg.  From Fig.\ \ref{fig:fgn},
it is readily seen that
\begin{eqnarray}
F^{gA}_{n,n+2j} - F^{gB}_{n,n+2j} & = & 0,  \nonumber \\
F^{gA}_{n,n+2j+1} - F^{gB}_{n,n+2j+1} & \equiv & -b^g_n,
\end{eqnarray}
where $j$ is a positive integer with the restriction $f^g_n$ and
$f^g_{n+2j}$ do not intersect, and $ b^g_n = b^g_{n+2}$ for any $n$.
Then, choosing $n=0$,
\begin{eqnarray}
D^g_{2j} - D^g_0 & = & 0,  \nonumber  \\
D^g_{2j+1} - D^g_0 & = & -b^g_0.
\end{eqnarray}
Again it follows that a SPO parameter $D^g \equiv D^g_0$ can be
associated to any VB configuration, so
\begin{equation}
D^g_n = D^g - \frac{1}{2}\left[ 1 - (-1)^n \right] b^g,
\label{eqn:Dn0g}
\end{equation}
with $b^g \equiv b^g_0$.

The general order parameter $D^g$ can be related to the previous one,
$D$, in a simple way.  For simplicity, without loss of generality, 
let us consider the fragment $F_{0,2j}$ limited by $f^g_0$ and
$f_{2j}$, $j > 0$.  We select $j$ in such a way that as $f^g_0$ and
$f_{2j}$ do not intersect.  $I_F$ is the number of SP with both sites
in $F_{0,2j}$;  $l_F$ ($r_F$) is the number of arrows connecting an
$A$ site in $R_{2j}$ ($L_0$) to a $B$ site in $L_0$ ($R_{2j}$), and
$F^{A}_{0,2j}$ ($F^{B}_{0,2j}$) is the number of sites belonging to
the intersection of $F_{0,2j}$ and sublattice $A$ ($B$). Then
\begin{eqnarray}
P^{g-}_0 - l_F + P^+_{2j} - r_F + I_F & = & F^{A}_{0,2j}, \nonumber \\
P^{g+}_0 - r_F + P^-_{2j} - l_F + I_F & = & F^{B}_{0,2j}.
\end{eqnarray}
Subtracting these two equations
\begin{equation}
-D^g + D_{2j} = F^{A}_{0,2j} - F^{B}_{0,2j}.
\end{equation}
Using Eq.\ (\ref{eqn:Dn0g}) we obtain
\begin{equation}
D^g = D - \left( F^A_{0,2j} - F^B_{0,2j} \right) ,
\end{equation}
which is independent of $j$, provided it is not too small to prevent
$f^g_0$ and $f_{2j}$ are intersecting.  Thence, there is a one to
one correspondence between allowed values of $D^g$ and $D$.

\subsection{LR-SPO of eigenstates and degeneracy}
\label{ssec:blocs}

So far, we have separated the model space in $w+1$ subspaces.
At this point we note that two singlets from different subspaces
must be different repeatedly at every position along the ladder.
Therefore, according to Pauling's island-counting technique
\cite{Pauling,Pauling-2}, they are asymptotically orthogonal
and non interacting via any interaction mediated by a few-particle
operator.  Then the matrix of the Hamiltonian asymptotically
block-diagonalises, so configurations belonging to different
subspaces do not mix in the configuration-interaction sense.  Thus
$D$ may be taken as a long-range order parameter labelling the
eigenstates $\Psi_D$ of the $D$ block.

Furthermore, ${\cal H}^w_D$ and ${\cal H}^w_{b-D}$ subspaces
are equivalent by a translation $T$ of one rung along the legs.
Then,
\begin{equation}
T \Psi_D = \Psi_{b-D}
\end{equation}
Consequently, degeneracy is always expected to occur, except for
$D = 0$ and $w$=even, i.e.\ $b=0$.

\section{Energy ordering}
\label{sec:counting}

Within the dimer-covering approximation the {\it resonance energy\/}
in units of $J$, $N \varepsilon_r(w,D)$, i.e.\, the ground-state
energy correction below the energy of a single Kekul\'e structure
(-0.375$N$), depends on the configuration interaction amongst the
different Kekul\'e structures.  It has been argued
\cite{Seitz,Zivkovic} that one might consider this energy lowering
to depend solely on the dimension of the space spanned by the
appropriate Kekul\'e structures.  Let $n(w,D)$ be the dimension of
the $D$ block for a $w$-legged ladder.  Since $n(w,D)$ is
multiplicative in terms of a break up into subsystems while the
energy is additive, such a functional dependence should be of the
form
\begin{equation}
\varepsilon_r(w,D) \approx -\frac{CJ}{wL} \ln n(w,D),
\label{eqn:E-r}
\end{equation}
where $C$ is a fitting parameter independent of the structure to
some degree, and in particular presumed to be (at least
asymptotically for large $w$) the same for all $w$-legged ladders.
For energy-ordering purposes the value of $C$ is irrelevant.
Nevertheless, for the nearest-neighbour isotropic Heisenberg model
the value of $C$ have been determined for a class of benzenoid
hydrocarbons \cite{Seitz} (with $C$=0.5667) and for finite
square-lattice fragments \cite{Zivkovic} (with $C$=0.75), by fitting
the logarithm of the Kekul\'e-structure count to the resonance energy
calculated from an equally-weighted Kekul\'e-structure wave function.

Arguments supporting Eq.\ (\ref{eqn:E-r}) relay on the fact that the
energy is an extensive magnitude, i.e.\ scales as the system size,
while the number of Kekul\'e structures needs to scale exponentially
with the system size.  Then, $C$ may be determined by fitting the
logarithm of the Kekul\'e-structure count to better estimates to the
energy than those used in Refs.\ \cite{Seitz} and \cite{Zivkovic}.
As a first attempt to do so, we obtain $C_w$=0.84, 1.0 and 0.93
appropriate for square-lattice strips with $L \rightarrow \infty$
and $w$=2, 3 and 4, respectively, when the ``exact'' energy values
of Table II in Ref.\ \cite{Zivkovic} have been used.  The weighted
average of them could be used as a rough estimate of $C$ appropriate
for the two-dimensional square lattice.  Assuming that the error of
$C_w$ typically behaves as $1/w$, we obtain $C = 0.94 \pm 0.19$,
although a more reliable value would be desirable.

The values of $n(w,D)$ can be easily obtained (see the Appendix) as
the $L/2$ power of the highest eigenvalue, $\Lambda^2_{wD}$, of the
$D$ block of the square of a dimer-covering-{\em counting\/} transfer
matrix ${\cal T}^2_D$.  Then, the resonance energy of the lowest
lying state in the subspace $D$ can be written as
\begin{equation}
\varepsilon_r(w,D) \approx -\frac{CJ}{w} \ln \Lambda_{wD}.
\label{eqn:ressonance}
\end{equation}

\section{Topological spin defects and LR-SPO}
\label{sec:defects}

Let us now suppose that there is a topological spin defect on site
$i$ of rung $n$, i.e.\ for any reason the site $ni$ remains
not spin-paired,
\begin{equation}
ni \in \left\{ \begin{array}{lr}
A, & {n+i}= {\rm  odd}, \\
B, & {n+i}= {\rm  even}.
\end{array} \right.
\end{equation}
Then, choosing boundaries parallel to rungs,
Eqs.\ (\ref{eqn:valans2}) become
\begin{eqnarray}
w^A_n = P^-_{n-1} -l + P^+_n -r +  I_n +
\frac{1}{2} \left[ 1-(-1)^{n+i}\right]  \nonumber  \\
w^B_n = P^+_{n-1} - r + P^-_n - l +  I_n +
\frac{1}{2} \left[ 1+(-1)^{n+i}\right]
\end{eqnarray}
and the recurrence relation across the rung $n$ with a topological
spin defect on site $ni$ is
\begin{equation}
D_n = D_{n-1} + (-1)^n \left[ b + (-1)^i \right] .
\label{eqn:rec-Ds3}
\end{equation}
If the order parameter to the left of the spin defect is $D_l$,
according to Eq.\ (\ref{eqn:DnD0}), the order parameter to the right
of site $ni$, $D_r$,
will be
\begin{equation}
D_r = D_l +(-1)^{n+i}.
\end{equation}
Then, a topological spin defect can be seen as a domain wall
separating sectors with order parameters $D_l$ and $D_r$.

Furthermore, since the subspaces $D$ and $b-D$ are degenerate, the
energy per site associated to the sectors to the right and to the
left of a spin defect located at the $ni$ site will be degenerate
if
\begin{equation}
D_l = \frac{1}{2} \left[ b - (-1)^{n+i} \right]
\end{equation}
This equation has a solution only when $b=1$, i.e.\ $w$ is odd, and
$D_l=0$ (with $D_r=1$) or $D_l=1$ (with $D_r=0$).

Furthermore, it is possible to form a local region between rungs $n$
and $m$ of LR-SPO $D \pm 1$ placing two spin defects in a wave
function of LR-SPO $D$, one in a site $ni$ and the other in a site
$mj$, with $n+i+m+j$=odd. 

\section{Results and discussion}
\label{sec:results}

Considering that $w$-legged (even) $L\rightarrow\infty$
antiferromagnetic spin-$\frac{1}{2}$ Heisenberg ladders are bipartite
systems with singlet ground state, we have assumed that their ground
state is described by an RVB many-body wave-function {\it
ans\"atze\/} within the space spanned by the $M$-range (covalent)
valence-bond configurations, ${\cal H}^w$, with arbitrarily large
$M\ll L$.

We have shown that any (covalent) VB configuration exhibits a LR-SPO.
This LR-SPO is characterised by a parameter $D$, which can take $w+1$
different values, and allows to specify the local (at boundary) array
of spin-pairings penetrating the boundary $n$, for any $n$.  The
shape of the boundary selected to define $D$ is quite arbitrary:  it
can go up and down, but it is self-avoiding and is not hitting any
site, so it is able to break up the ladder in two (non-longitudinal)
parts.  When the boundary is chosen to be parallel to the rungs, the
allowed values of $D$ are: 
\begin{equation}
D = \frac{w+b}{2}, \frac{w+b-2}{2}, \dots , \frac{b-w}{2},
\end{equation}
where $b$ is zero (one) for $w$=even (odd).  Choosing another 
boundary, the values of $D$ simply change by a boundary-dependent
integer.

This long-range spin-pairing-order allows to separate ${\cal H}^w$
in $w+1$ subspaces, ${\cal H}^w_D$.  Two configurations in different
${\cal H}^w_D$ differ repeatedly on each of the L rungs of the
ladder.  Consequently, they are asymptotically orthogonal and non
interacting via any interaction mediated by a few-particle operator.
Thence, the matrix of the Hamiltonian asymptotically
block-diagonalises.  Thus, the eigenstates of $H$ do not mix
configurations with different value of $D$.  Therefore, $D$ may be
taken to be a long-range order parameter labelling the eigenstates of
the $D$ block.  This gives rise to $w+1$ RVB {\it Anz\"atze\/}
exhibiting different LR-SPO, $\Psi_D$.  It is also worth noting that
asymptotic strong orthogonality and LR-SPO in RVB wave functions
persist with a long finite cut-off.

Furthermore, subspaces ${\cal H}^w_D$ and ${\cal H}^w_{b-D}$ are
equivalent by a translation of one rung along the legs, $T$\@.  Then
\begin{equation}
T \Psi_D = \Psi_{b-D},
\end{equation}
and the energy of the corresponding wave functions must be
\begin{equation}
E_D = E_{b-D}.
\label{eqn:degeneracio}
\end{equation}
Thus degeneracy is always predicted to occur, except for $w$=even and
$D=0$, irrespective of the details of any Hamiltonian preserving
translational symmetry.

We have obtained the values of $\Lambda_{wD}$ of
Sec.\ \ref{sec:counting} for $w$ ranging from one to twelve and $D$
from $b$ to $(w+b)/2$.  Table \ref{tab:fases} summarises the zero
order resonance energies $\varepsilon_r(w,D)$, in units of $C$,
calculated by using Eq.\ (\ref{eqn:ressonance}).  Since it is
unlikely that the zero-order energies are drastically modified by
the small corrections to the isotropic nearest-neighbour Heisenberg
Hamiltonian, we expect an energy ordering
\begin{equation}
E_b < E_{b+1} < \cdots < E_{\frac{1}{2}(w+b)}.
\label{ordenacio}
\end{equation}
Therefore, for $w$=even, the ground state belongs to the
non-degenerate $D=0$ subspace, with
\begin{equation}
T \Psi_0 = \Psi_0,
\end{equation}
while, for $w$=odd, the ground-state manifold is spanned by $\Psi_0$
and $\Psi_1$, which are eigenstates of $H$, but not of $T$\@.  The
eigenstates $\Phi_\pm$ of the translation operator, are defined in
the ground-state manifold,
\begin{equation}
\Phi_\pm \equiv \frac{1}{\sqrt{2}} \left( \Psi_0 \pm \Psi_1 \right) ,
\end{equation}
so
\begin{equation}
T \Phi_\pm = \pm \Phi_\pm .
\end{equation}
with $k=0$ and $k=\pi$.  This RVB asymptotic degeneracy for $w$=odd
is consistent with a very wide body of evidence.  

It could be argued that the true ground state of the
spin-$\frac{1}{2}$ nearest-neighbour Heisenberg Hamiltonian for
$w=1$ and $L \rightarrow \infty$ should be described in terms of two
degenerate subspaces, in order to regain the non-degenerate and
gapless Bethe {\it Ansatz\/} state.  For finite (even) $L$ with
cyclic boundary conditions the two degenerate RVB wave functions do
interact, in the configuration interaction sense, giving a splitting
of the translationally-adapted states $\Phi_\pm$ (with $k=0$ and
$k=\pi$), which leads to a non-degenerate ground state.  Thus,
for finite $L$, restriction to just one subspace does not yield a
true ground state.  Nevertheless, when the splitting is examined as
a function of $L$, it is seen that the gap closes at least as fast
as $1/L$ and the states $\Phi_\pm$ approach to an accidental
degeneracy.  This feature is basically related with the dimerzing
instability of the Bethe {\it ansatz\/} state.  For instance, if a
small dimerizing interaction occurs, then the accidental degeneracy
of Eq.\ (\ref{eqn:degeneracio}) no longer holds.  Therefore, either
$\Psi_0$ or $\Psi_1$ will be lower in energy and completely dominate
the wave function, leading to a spin-Peierls broken-symmetry ground
state \cite{PRB-79}.

The $w \rightarrow \infty$ limit is obtained by fitting
$\varepsilon_r(w,0)/C$ by a power series in $1/w$,
\begin{equation}
\frac{\varepsilon_r(w,0)}{C} \approx
\frac{\varepsilon_r(\infty ,0)}{C} +\frac{a_1}{w} +\frac{a_2}{w^2}.
\label{eqn:fit} 
\end{equation}
In Table \ref{tab:fit} the values of $\varepsilon_r(\infty ,0)/C$,
$a_1$ and $a_2$ are collected for both $w$=even and $w$=odd, along
with their weighted average for $\varepsilon_r(\infty ,0)/C$.
For the purpose of seeing how reasonable this simple model is,
a value of $C$ is needed.  When $C$ is fitted to dimer-covering
estimates of the energy for square-lattice fragments, the value 
$0.75$ is obtained \cite{Zivkovic}.  Using this value of $C$, the
present simple model predicts an energy per site of $\approx -0.594$
(in unit of $J$), in good agreement with the value $-0.604$
calculated at the same level by Liang, Doucot and Anderson
\cite{Liang}.  When $C$ is evaluated as the weighted average of
the $C_w$, $w$=2, 3 and 4, derived from the ``exact'' estimates of
the energy of Zivkovic {\em et al.\/} \cite{Zivkovic} (see section
\ref{sec:counting}) we obtain $C = 0.94 \pm 0.19$.  Obviously, a more
reliable $C$ value would be desirable.  Using this value of $C$, an
energy per site of $\approx -0.65 \pm 0.10$ (in unit of $J$) is
obtained, which also compares fairly well with the best estimate
($-0.668$) of Liang, Doucot and Anderson \cite{Liang}.

We have shown that a topological spin defect, i.e.\ a non-paired
site, can be seen as a domain wall separating two sectors of the
ladder with order parameters $D_l$ and $D_r$ 
\begin{equation}
D_r = D_l +(-1)^{n+i},
\end{equation}
Furthermore, the energy per site associated to the right and to the
left sectors of the spin defect will be degenerate only when $b=1$,
i.e.\ $w$ is odd, and $D_l=0$ (with $D_r=1$) or $D_l=1$
(with $D_r=0$).  Also, it is possible to form a local region between
rungs $n$ and $m$ of LR-SPO $D \pm 1$ placing two spin defects in a
wave function of LR-SPO $D$, one in a site $ni$ and the other in a
site $mj$, with $n+i+m+j$=odd. 

Now, half-filled excited states (other than the $d_n - 1$ singlets
orthogonal to the ground state) or even slightly doped states are
analysed via topological spin defects.  There are different types of
excitations conceivable from a {\it Maximally-spin-paired\/} ground
state.  For instance, preserving half filling (one electron per
site), there are primarily spin excitations.  In this case, two
topological spin defects, one in an $A$ site and the other in a $B$
site, are obtained by breaking one SP to form a triplet state.
Away from half filling, removing (adding) one electron produces a
vacant (doubly-occupied) site and the ladder becomes a doublet, with
two sites, again one in an $A$ site and the other in a $B$ site,
that cannot be SP\@.  In this case hopping terms must be retained in
the Hamiltonian and the so-called {\it t-J model\/} applies.
Thence, the doublet is a weighted superposition of configurations
with a single spin not SP and a vacant (doubly-occupied) site.
Either a vacancy or a doubly-occupied site may also be assimilated
to a topological spin defect, although there is no spin associated
with them.  Then, away from half filling, it may be conceivable a
local region limited by two vacant (doubly-occupied) sites, or even
a vacant (doubly-occupied) site and a single non-SP spin (provided
that the doping is not so strong as to preclude a
maximally-spin-paired ground state).  Therefore, there are
low-energy spin and charge excitations.  Still, going up in the
hierarchy of Hamiltonians, the Hubbard or even a more general
Hamiltonian has to be considered.  In this case, still another type
of excitations (though presumably of higher energy if a
Heisenberg-like Hamiltonian is assumed to govern the lowest-lying
region of the spectrum) can be produced relaxing the
single-occupancy constrain.  This leads to the {\it ionic\/} states,
i.e.\ states with at least a couple of sites, one doubly occupied
and the other empty.  Therefore, the couple of spin defects,
associated to an excitation above a {\it Maximally spin-paired\/}
state $\Psi_D$, are limiting a local region of LR-SPO $D \pm 1$.

When $w$=odd and $D$=0 ($D$=1), there can be local regions with
$D=\pm 1$ ($D$=0, 2).  Then, since subspaces $D$=0 and $D$=1 are
degenerate, it is possible to have a local region with identical per
site energy inside and outside the local region.  In this case the
topological spin defects limiting the local region are not confined,
though it may happen that they attract one another (with an ordinary
short-range potential).  Two conclusions can be drawn from this
result.  First, for the half filling case, triplets with the two
spin defects very far apart from one another are possible.  Although
breaking a singlet does cost some energy (due to contribution of the
diagonal terms), there is a gain in kinetic energy (off-diagonal
terms contributions), since (for $w$=odd) the two spins are not
confined and can move independently.  Consequently, a gapless
triplet spectrum is not inconsistent with the results of this work.
This feature can be understood as a generalisation to any odd-legged
spin-$\frac{1}{2}$ antiferromagnetic ladders of Lieb, Shultz, and
Mattis theorem \cite{LSM,Yamanaka} holding for one-dimensional
systems.  Second, away from half filling, removing (adding) an
electron to the system yields a {\em non-confined\/} pair of sites,
one being a vacant (doubly-occupied) site and the other a non-SP
site.  It is worth noting that the vacant (doubly-occupied) site
holds the charge, while the non-SP site holds the up or down spin,
leading to charge-spin separation.

In clear contrast, when $w$=even, the order parameter of the local
region limited by the couple of spin defects is always associated to
higher energy per site.  This indicates that the couple of spin
defects should remain as close as possible so confinement is
predicted to occur.  Hence, at half filling, the energy difference
between the triplet and the ground state will be finite, and the
$w$=even ladders are expected to be gapped.  Nevertheless, we
observe that the energy difference per site between the lowest lying
subspaces, $\Delta\varepsilon$, behaves as $\sim w^{-\alpha}$, with
$\alpha \approx 1.8$, i.e.\ $\Delta\varepsilon$ decreases faster
than $1/w$.  Therefore, a lowering of confinement and the closing of
the gap is predicted for increasing values of $w$.  At this point,
it could be argued that an excitation energy lowering can be
achieved by allowing a Bloch superposition of analogous couples of
topological spin defects.  However this lowering is not expected to
be as important as to close the gap, at least for $w$=2
\cite{Zivkovic}.  Nevertheles, within the scope of the present work
it is not able to predict the energy ordering of the lowest singlet
and triplet excited states.  Now, away from half filling, when
removing (adding) one electron the vacant (doubly-occupied) site will
be bound to the non-SP spin, forming a ``quasi-particle'' with charge
and spin.   Again we expect the lowering of the confinement and the
charge-spin separation with increasing values of $w$.  This
quasi-particle is far from being a static vacant (doubly-occupied)
site bound to a static non-SP site.  The hopping term of the
{\it t-J\/} Hamiltonian allows the vacant (doubly-occupied) site to
move while the exchange part mixes up all the doublet VB
configurations with the constrain of keeping the vacant
(doubly-occupied) site and the non-SP spin close to one another.
Therefore, the appropriate wave function must be a weighted
superposition of all possible static configurations in order to
benefit from the hopping contribution to the energy, while lowering
the exchange energy by keeping both the spin and the vacant
(doubly-occupied) site bound.  When removing (adding) two electrons,
arguments based on the LR-SPO alone are not able to decide if the
two-holes (two-electrons) state is described as two quasi-particles
or two bound vacant (doubly-occupied) sites.  Nevertheless, we expect
the two vacant (doubly-occupied) sites to be confined to benefit from
the energy lowering due to the hoping term of the Hamiltonian, as
has been pointed out in Ref.\ \cite{White-97}, using numerical
results from DMRG techniques on clusters.  Another argument to
take into account is the range of the RVB\@.  Since the two
quasi-particles are expected to couple to a singlet, non-bound
quasi-particle would imply a long bond, while a short-range RVB is
expected for $w$=even ladders \cite{PRB-91,White}.

The above results also apply to open boundary conditions along the
$L$ direction.  In this case, no pairing is expected to the left of
rung $n=1$, and to the right of rung $n=L$\@.  Then, open boundary
conditions determine the long-range order parameter to be $D=0$, with a
pair of spin defects still delimiting a $D=1$ fragment, which is
degenerate with that of $D=0$ solely for $w$=odd, so results above
still apply.  Furthermore, if the boundary conditions are so that
would force an order parameter of energy $\varepsilon_r(w,D)$
($\varepsilon_r(w,b-D)$), $D\neq b$, we expect that a state with
$D-b+1$ spin defects located close to each end of the ladder, these
defects limiting an extended area of resonance energy per site
$\varepsilon_r(w,b)$, would be favoured.

The results of the present work relay on two conditions.  First,
the bipartitioning condition, which allows to establish the LR-SPO.
Second, the energy ordering given by Eq.\ \ref{ordenacio}.  This
energy ordering is valid when the dominant part of the Hamiltonian
is the isotropic sin-1/2 Heisenberg Hamiltonian.  It might happen
that the bipartitioning condition is not fulfilled and/or the
corrections away from the isotropic spin-1/2 Heisenberg Hamiltonian
are as important as the energy ordering is not that of
Eq.\ (\ref{ordenacio}), then the above discussion will not apply.

\section{Conclusions}
\label{sec:conclusions}

Identifying a LR-SPO and understanding energy ordering and degeneracy
among RVB wave functions, as well as the discontinuities in the
LR-SPO introduced by topological spin defects, allow a general and
systematic explanation of the differences between even and odd
ladders.  It provides an intuitive understanding of the physics of
spin ladders in general.  In particular, for odd leg ladders
\begin{enumerate}
\item a doubly degenerate ground state, breaking translational
symmetry, is obtained;
\item a gapless triplet spectrum is consistent with the results of
the present work;
\item  a simple understanding of the deconfinement of spin
excitations as well as charge-spin separation is provided.
\end{enumerate}
On the contrary, for small $w$=even ladders
\begin{enumerate}
\item a non degenerate ground state is obtained;
\item  triplet spin excitations are found to be gaped;
\item  a simple understanding of the confinement of spin excitations
is provided;
\item  a lowering of the confinement and the closing of the gap is
predicted for increasing values of $w$.
\end{enumerate}

\begin{acknowledgement}

The author acknowledges valuable discussions with D. J. Klein.  This
research was supported by the DGICYT (project PB95-0884).

\end{acknowledgement}

\appendix
\section*{Dimer covering counting}
\label{sec:appendix}

Let us analyse from a local point of view the dimer-cov\-er\-ing
singlets.  We can identify a dimer-covering local state,
$\mid e_{nI} )$, $I$ ranging, according to which legs have an arrow
across the $f_n$ boundary.  The direction of any arrow is fixed by
$n$ and the leg number.  Then, it can be seen that there
are $2^w$ different local states for each boundary, which can be
classified according to the value of $D_n$, $\mid e^D_{nI} )$.
The local states of position $n+1$ are mirror images of those of
position $n$.

A dimer-covering-{\em counting\/} matrix, ${\cal T}_n$, is defined as
\begin{equation}
(e_{n-1 I}\mid {\cal T}_n \mid e_{n J}) = \left\{ \begin{array}{ll}
1, & \mid e_{n J}) \textrm{ can succeed } \mid e_{n-1 I}) \\
0, & \textrm{otherwise}
\end{array} \right.
\end{equation}
Then, the number of dimer-covering states in a ${\cal H}^w_D$
subspace is
\begin{equation}
n(w,D) = \sum_{e^D_{0I}}
( e^D_{0I}\mid {\cal T}_1{\cal T}_2 \dots {\cal T}_L \mid e^D_{0I}).
\end{equation}
Since for any dimer-covering singlet $D_{n-1} = D_{n+1}$, ${\cal T}_n
{\cal T}_{n+1}$ is a block-diagonal symmetric matrix that does not
depend on $n$ (apart from the direction of the arrows in the local
states that it relates) we can omit the sub-index.  For
$L\rightarrow\infty$, the highest eigenvalue $\Lambda^2_{wD}$ of the
$D$ block ${\cal T}^2_D$ dominates, and
\begin{equation}
n(w,D) \approx \Lambda^L_{wD}.
\end{equation}

\section*{Figure captions}
\label{sec:figures}

\begin{figure}[h]
\caption{A complete set of linearly independent (covalent)
singlets for a six-sites system:  a) $1 \times 6$;
b) $2 \times 3$.}
\label{fig:sing}
\end{figure}

\begin{figure}[h]
\caption{A fragment of a $w$=9 ladder, showing (inside a circle) the
value of $D_n$ at each boundary $f_n$.  Note that
$D_n = D_{n-1}- (-1)^nb$ (sublattices $A$ and $B$ are selected as
$w^A_n - w^B_n = (-1)^n b$).}
\label{fig:LRO-K}
\end{figure}

\begin{figure}[h]
\caption{A dimer-covering configuration is an eigenstate of
${\bf S}_j \cdot {\bf S}_k$ when sites $j$ and $k$ are spin-paired.
The ``off-diagonal'' singlet with $(i,l^*)$ and $(k,j^*)$ pairings is
also obtained when ${\bf S}_j \cdot {\bf S}_k$ acts on a VB
configuration with spin-pairings $(i,j^*)$ and $(k,l^*)$.}
\label{fig:recoup}
\end{figure}

\begin{figure}[h]
\caption{A fragment of a $w$=7 ladder, showing (inside a circle) the
value of $D_n$ at each boundary $f_n$.  Note that still
$D_n = D_{n-1}+(-1)^n b$ (sublattices $A$ and $B$ are selected as
$w^A_n - w^B_n = (-1)^n b$), although here there are longer-range
spin-pairings.}
\label{fig:LRO}
\end{figure}

\begin{figure}[h]
\caption{Portion of a ladder showing boundaries $f^g_n$ and $f^g_m$
which go up and down:  a) $m-n$=even;  b) $m-n$=odd.  These
non-intersecting boundaries define a fragment of the ladder (the set
of sites with a full dot or star), $F^g_{n,m}$, as the intersection
of $R_n$ and $L_m$.  Analysing the fragment a leg at time, it is
worth noting a series of facts:  First, there are $m-n$ sites in each
leg.  Also, whenever a site $qi$ belongs to $F^g_{n,m}$, neither the
site $T^{m-n}qi$ nor the site $T^{-m+n}qi$ ($qi$ translated $m-n$
steps to the right or to the left) are included in the fragment.
Furthermore, in (a), $T^{m-n}qi$ remains in the same sublattice than
$qi$ while, on the contrary, in (b), $T^{m-n}qi$ belongs to a
different sublattice than $qi$.  Finally and more important, in (a)
half of the $m-n$ sites belong to the sublattice $A$ and the other
belong to the $B$ sublattice, while in (b) the fragment has different
number of sites in the sublattice $A$ than in the sublattice $B$.}
\label{fig:fgn}
\end{figure}

\newpage
\section*{Table captions}
\label{sec:taules}

\begin{table}[h]
\caption{The absolute value of the resonance energy in units of
$C$, $(\ln \Lambda_D)/w$, for the lowest lying state of subspaces with
order parameters from $D=0$ to $(w+b)/2$, and the extrapolations to
$w \rightarrow \infty$ for the lowest lying subspace of both, the
$w$=even and $w$=odd, series.\protect\label{tab:fases}}
\end{table}

\begin{table}[h]
\caption{Parameters of the Eq.\ (\protect\ref{eqn:fit}), which
fits the ground-state ($D=0$) resonance energy, in units of $C$,
for the $w$=even and $w$=odd series and their weighted average.
\protect\label{tab:fit}}
\end{table}

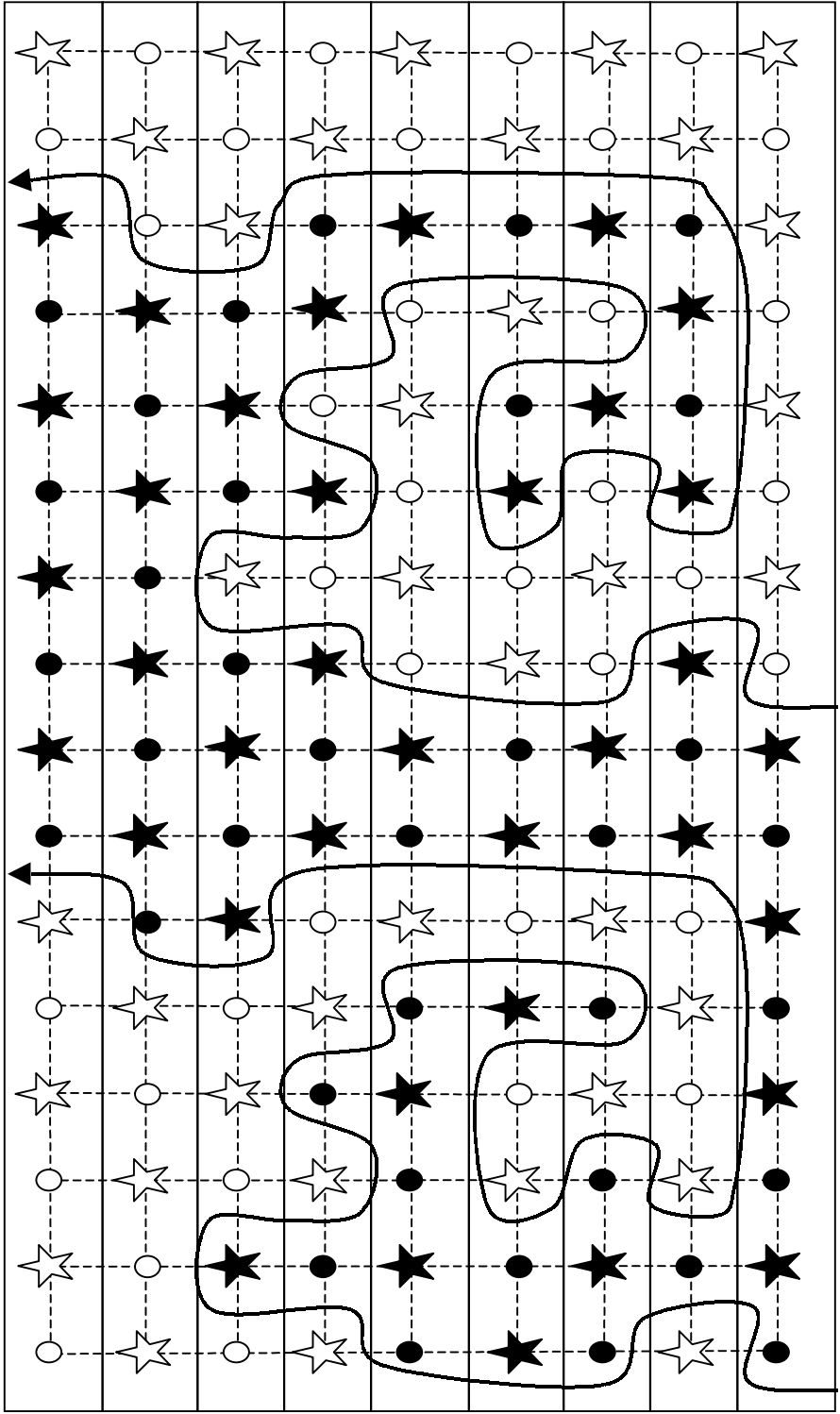
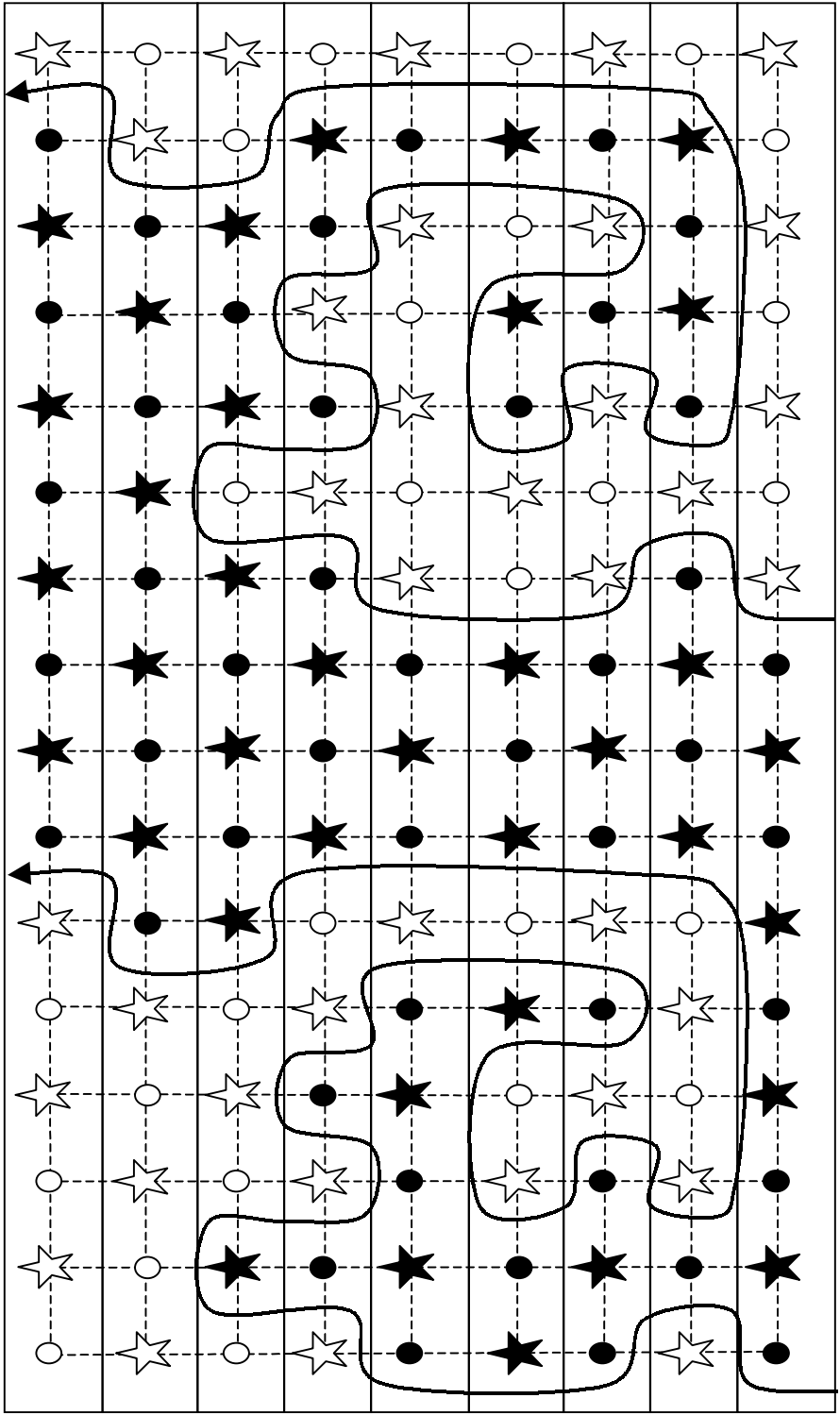
\end{document}